\documentclass[twocolumn,longbibliography,
superscriptaddress,
amsmath,amssymb,aps,pra]{revtex4-2}

\usepackage{hyperref}

\usepackage{graphicx}
\usepackage{dcolumn}
\usepackage{bm}

\usepackage{ulem}
\usepackage{color}

\newcommand{\wzy}[1]{{\color{black} #1}}

\begin{document}
	\title{Thouless pumping via discrete jumping of the adiabatic Hamiltonian}%
	
	\author{Lang Yu}
	\affiliation{Key Laboratory of Atomic and Subatomic Structure and Quantum Control (Ministry of Education), Guangdong Basic Research Center of Excellence for Structure and Fundamental Interactions of Matter, School of Physics, South China Normal University, Guangzhou 510006, China} 
	
	\author{Yihan Wu}
	\affiliation{Key Laboratory of Atomic and Subatomic Structure and Quantum Control (Ministry of Education), Guangdong Basic Research Center of Excellence for Structure and Fundamental Interactions of Matter, School of Physics, South China Normal University, Guangzhou 510006, China} 
	
	\author{Zhen-Yu Wang}
	\email{zhenyu.wang@m.scnu.edu.cn}
	\affiliation{Key Laboratory of Atomic and Subatomic Structure and Quantum Control (Ministry of Education), Guangdong Basic Research Center of Excellence for Structure and Fundamental Interactions of Matter, School of Physics, South China Normal University, Guangzhou 510006, China} 
	\affiliation{Guangdong Provincial Key Laboratory of Quantum Engineering and Quantum Materials, Guangdong-Hong Kong Joint Laboratory of Quantum Matter, South China Normal University, Guangzhou 510006, China}

	\begin{abstract}
		Thouless pumping is a celebrated topological effect that enables quantized particle transport and holds promise for applications in quantum technologies. However, the standard adiabatic pumping theory dictate that the Hamiltonian must be varied continuously and slowly throughout the whole pumping process. In this work, we show that Thouless pumping can be realized by using only discrete sampling of the Hamiltonian, removing the conventional constraint of continuous and slow change of control. Specifically, we select particular parameter points of the Hamiltonian for conventional continuous quantum adiabatic evolution, and then apply the Hamiltonian at these parameter points which induces a geodesic evolution for the pseudospin with a given quasimomentum. This realizes an adiabatic evolution for the Thouless pumping according to a necessary and sufficient condition for quantum adiabatic evolution. 
	\end{abstract}
	
	\maketitle
	
	\section{Introduction}
	Thouless pumping has been widely studied as a fundamental phenomenon of topological transport in condensed matter and atomic physics~\cite{Thouless1983,Niu1984,Xiao2010,Citro2023,Nakajima2016,Lohse2016,Lohse2018}. Originally introduced by Thouless, it demonstrates that adiabatic cyclic modulation of a one-dimensional periodic potential can yield quantized particle transport per cycle~\cite{Thouless1983}. Characterized by an Chern number~\cite{Xiao2010,Citro2023}, Thouless pumping serves as a dynamical analog of the integer quantum Hall effect~\cite{Citro2023}, rendering the quantized particle transport intrinsically robust against weak disorder and interactions~\cite{Niu1984}. Over the past decade, Thouless pumping has been experimentally demonstrated across a variety of platforms, including ultracold atom~\cite{Nakajima2016,Lohse2016,Lohse2018} and photonic systems~\cite{Benalcazar2022,Kraus2012,Zilberberg2018,Cerjan2020}.
	
	In the conventional approach of Thouless pumping, the Hamiltonian is changed continuously and slowly according to the conventional condition for quantum adiabatic processes~\cite{Messiah1965}. This slow continuous change and hence long evolution time impose challenges for the experimental implementation of Thouless pumping~\cite{Privitera2018,Arceci2020,Fedorova2020,Schouten2021,Malikis2022}.
	On the possibility to achieve Thouless pumping beyond the conventional adiabatic processes, a recent study~\cite{Liu2025} proposed the use of shortcuts to adiabaticity (STA)~\cite{Berry2009,
		Demirplak2005,GueryOdelin2019,An2016,Chen2010a,Chen2010b} to speed up Thouless pumping. However, this STA scheme relies on highly intricate auxiliary control fields, making it difficult to implement experimentally.
	
	It was found that the conventional condition for quantum adiabatic processes can be problematic~\cite{Wang2016}. The development of a necessary and sufficient condition for quantum adiabatic evolution~\cite{Wang2016} indicated that it is possible to employ piecewise-constant Hamiltonian for the suppression of non-adiabatic transitions~\cite{Wang2016,Xu2019,Chen2024,Zhang2025,Zheng2022,Gong2023,Xing2026}. With experimental verification~\cite{Xu2019}, the key mechanism for quantum adiabatic evolution is not the slow variation of the Hamiltonian compared with the minimal energy gap but the fast averaging by the dynamic phase factors~\cite{Wang2016}. Remarkably, this idea can be used to break the quantum adiabatic speed limit by jumping along a geodesic~\cite{Xu2019}. 
	Based on the new condition, accelerated quantum adiabatic schemes have been developed for qubits~\cite{Xu2019,Zheng2022,Zhang2025}, three-level systems~\cite{Gong2023}, multisqueezed states~\cite{Chen2024}, and for more general multi-level quantum systems~\cite{LiuWang2022,Xing2026}. Although the theory of the necessary and sufficient condition for quantum adiabatic evolution has been applied to these simple systems, its application in condensed matter physics remains unexplored. 
	
	In this work, we show that one can modify the conventional adiabatic control for Thouless pumping to a fundamentally new form based on the necessary and sufficient condition for quantum adiabatic evolution. That is, instead of using the continuous and slow varying Hamiltonian required in the traditional adiabatic condition, we only apply the Hamiltonian at a finite number of discrete parameter points of the parameter space used for traditional Thouless pumping. This allows the suppression of nonadiabatic transitions for Thouless pumping in a robust manner and would simplify the complexity of experiment because only a much smaller subset of control parameter space is used compared with the traditional Thouless pumping. Compared with the STA technique, our method is inherently adiabatic and does not need to apply additional complex auxiliary fields, providing a route to manipulate complex quantum systems without sophisticated control. 
	
	This paper is organized as follows. In Sec.~\ref{sec:Theory}, we develop the Thouless pumping using discrete jumping of the adiabatic Hamiltonian. We then numerically study the jumping approach in momentum space and position space in Sec.~\ref{sec:Numerical}. We draw our conclusion in Sec.~\ref{sec:Conclusion}.

	\section{Theory for Thouless pumping using adiabatic jumping\label{sec:Theory}}
	
	\subsection{Thouless pumping}
\begin{figure}
		\centering
		\includegraphics[width=0.5\textwidth]{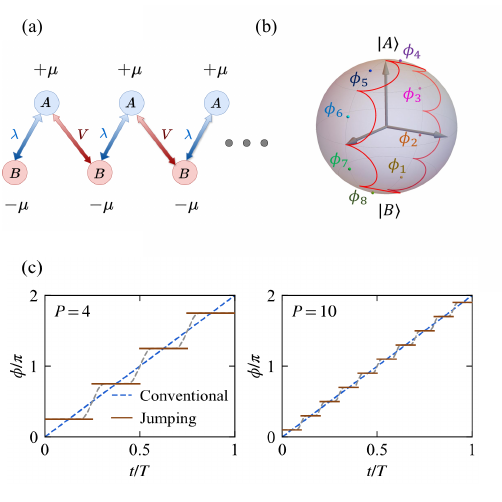}
		\caption{(a) Schematic of the Rice-Mele model with periodically modulated parameters $\mu$, $\lambda$, and $V$ for Thouless pumping. (b) Trajectory (red curve) for the $k=l$ pseudospin under the adiabatic jumping control with $P=8$. The rotating axes for the trajectory are determined by $H(\phi)$ with $\phi= \phi_1, \phi_2, \ldots, \phi_P$. (c) The functional forms of $\phi(t)$ for the conventional Thouless pumping (blue dashed lines) and adiabatic jumping control (stepwise red solid line) in one pumping period $T$ for different $P$. \wzy{The gray dashed line shows the smoothed version of the jumping scheme, where each abrupt jump is replaced by a sinusoidal transition [see Eq.~\eqref{sinusoidaltransition}].}}
		\label{fig:schematic}
	\end{figure}

	We begin by reviewing the conventional Thouless pumping in the Rice–Mele model~\cite{RiceMele1982}. The Hamiltonian of the model reads~\cite{Asboth2016}
	\begin{multline}
		H = \sum_{m=-l}^{l} \big( V |A,m\rangle \langle B,m| + \lambda |B,m\rangle \langle A,m+1| + \mathrm{H.c.} \big) \\
		+ \mu |A,m\rangle \langle A,m| - \mu |B,m\rangle \langle B,m|,
	\end{multline}
	where $V = V(t)$ is the hopping strength between sites $A$ and $B$ within the same unit cell labeled by $m$, $\lambda = \lambda(t)$ is the hopping strength between different unit cells, and $\mu = \mu(t)$ determines the on-site potential energies of sites $A$ and $B$. See Fig.~\ref{fig:schematic}(a) for a sketch of the model. For convenience we choose the coordinate of the unit cells such that the lattice is centered at the origin, by defining $l=(N-1)/2$ for a total number $N$ of unit cells. Throughout this work, we adopt units in which $\hbar=1$. 
	
	To achieve Thouless pumping, one periodically modulates the parameters as~\cite{Liu2025}
	\begin{align}\label{eq:parament}
		V(t) & = J + \delta_0 \sin\phi(t), \\ \nonumber
		\lambda(t) & = J - \delta_0 \sin\phi(t), \\ \nonumber
		\mu(t) & = \Delta_0 \cos\phi(t),
	\end{align}
	where the phase $\phi=\phi(t)$ denotes the functional form for the periodic modulation with a period $T$. For conventional Thouless pumping, the control parameters are changed continuously, e.g., using a continuous change of the phase $\phi(t) = \phi_0 + \omega t$, where $\phi_{0}$ is the initial phase and $\omega$ is the modulation frequency with $T = 2\pi/\omega$ being the pumping period. For simplicity, we set $\phi_{0}=0$. Later in our jumping approach for Thouless pumping, we use a different functional form of $\phi(t)$. Here the parameter $J$ denotes the average hopping strength between neighboring sites, while $\delta_0$ and $\Delta_0$ are the modulation amplitudes of the hopping and on-site energy, respectively.
	
	We assume the periodic boundary condition $|m+N\rangle = |m\rangle$ for the basis states $\{|m\rangle\}$ in real space. Then the Fourier transform
	\begin{equation}\label{eq:Fourier_transform}
		|k\rangle = \sum_{m=-l}^{l} \frac{1}{\sqrt{N}} e^{i \frac{2\pi}{N} m k} |m\rangle,
	\end{equation}	
	defines the orthonormal momentum basis $\{|k\rangle\}$ with $k\in\{-l,-l+1,\cdots,l\}$  corresponding to $N$ discrete points in the first Brillouin zone. See Appendix~\ref{app:displacement} for the relation of the operators in the position $\{|m\rangle\}$ and momentum $\{|k\rangle\}$ bases. This transforms the Hamiltonian to a block diagonalized form:   
	\begin{equation}\label{eq:momentum_basis}
		H= H(\phi) = \sum_{k=-l}^{l} H_k(\phi) \otimes |k\rangle\langle k|,
	\end{equation}
	where we have emphasized that the Hamiltonian depends on the functional form of the phase $\phi=\phi(t)$ [see Eq.~\eqref{eq:parament}]. Here  
	\begin{equation}\label{eq:pseudospin Hamiltonian}
		H_k(\phi) = h_x^{(k)}(\phi) \sigma_x + h_y^{(k)}(\phi) \sigma_y + h_z^{(k)}(\phi) \sigma_z,
	\end{equation}
	is pseudospin Hamiltonian in the $k$-space with $\sigma_{\alpha}$ ($\alpha = x,y,z$) being the Pauli operators and the coefficients $h_x^{(k)} = V + \lambda \cos\left(\frac{2\pi }{N}k\right)$, $h_y^{(k)}= \lambda \sin\left(\frac{2\pi }{N}k\right)$,  $h_z^{(k)} = \mu$. Regarding $|A\rangle$ ($|B\rangle$) as the spin up (down) state of the pseudospin, we write $\sigma_{z}=|A\rangle\langle A|-|B\rangle\langle B|$ and  $\sigma_{x}=|A\rangle\langle B|+|B\rangle\langle A|$.

	In topological Thouless pumping, the quantized particle transport per cycle is characterized by the Chern number $C$, which can be expressed as the integral of the Berry curvature over the parameter space~\cite{Xiao2010}. We are interested in the quantity of a cyclic evolution
	\begin{equation}
		\bar{y}(T) = -\frac{d}{2\pi} \int_{\text{BZ}} dk \int_0^{T} dt \mathcal{B}(k,t),
	\end{equation}
	where BZ denotes the first Brillouin zone, $T$ is the pumping period of a cyclic evolution, $d$ is the length of each unit cell, and
	\begin{equation}
		\mathcal{B}(k, t) = \frac{1}{2}[\partial_k \vec{\sigma}(k, t) \times \partial_t \vec{\sigma}(k, t)] \cdot \vec{\sigma}(k, t),
	\end{equation}	
	with $\vec{\sigma} (k, t) = -\langle\psi(k,t)|\boldsymbol{\sigma}|\psi(k,t)\rangle$ being the Bloch vector, and $\boldsymbol{\sigma}=(\sigma_{x},\sigma_{y},\sigma_{z})$ following the notations in Ref.~\cite{Liu2025}. The time-evolved state $|\psi(k,t)\rangle=U_k(t)|\psi(k,0)\rangle$, where $|\psi(k,0)\rangle$ is the ground state of $H_{k}$ at $t = 0$ for each quasimomentum mode $k$ and the evolution operator
	\begin{equation}
		U_k(t)=\mathcal{T}\exp\left(-i\int_{0}^{t} H_{k}dt'\right) \label{eq:U_k_General}
	\end{equation}
	with $\mathcal{T}$ being the time-ordering operator. 
In an ideal adiabatic process, the evolved states $|\psi(k,t)\rangle$ follow exactly the Hamiltonian instantaneous eigenstates and $\bar{y}(T)$ gives the Chern number, that is $\bar{y}(T)/d = C$.  For non-adiabatic evolution, $\bar{y}(T)/d$ in general is different from  $C$. Therefore, the quantity  $\bar{y}(T)$ characterize the adiabaticity of the pumping process. 

In the conventional Thouless pumping, which uses the traditional adiabatic condition~\cite{Messiah1965} for the implementation of adiabatic evolution, the parameters of the Hamiltonian must be changed continuously and slowly. Therefore, the phase $\phi(t) = \omega t$ for the change of the Hamiltonian in the conventional Thouless pumping is varied in time continuously, see Fig.~\ref{fig:schematic}(c). When $\omega$ is much smaller than the minimum energy gap of the Hamiltonian through out the whole process of system evolution, the state $|\psi(k,t)\rangle$ evolves adiabatically. And as a consequence, $\bar{y}(T)/d \approx C$, ensuring that the wave packet shifts by exactly one unit cell at the end of each pumping cycle, as we will numerically demonstrate in Sec.~\ref{sec:Numerical}. 

\subsection{Thouless pumping via adiabatic jumping}
It is possible to remove the constraint of traversing continuously over all the parameter space by using the necessary and sufficient condition for quantum adiabatic evolution~\cite{Wang2016}.  According to~\cite{Wang2016}, the fundamental principle underlying adiabatic evolution is the cancellation of nonadiabatic transitions through the averaging by the dynamic phase factors over the evolution path.  When the trajectory of the parameters is sampled such that the dynamic phase factors are fast oscillating with a zero mean over the intervals of system evolution, the effect of non-adiabatic transitions are averaged out. This allows to implement a new kind of Thouless pumping even the parameters are not continuously varied. As we will demonstrate in Sec.~\ref{sec:Numerical}, indeed we can realize quantum adiabatic process for the realization of Thouless pumping even when only a smaller subset of the parameters for the Hamiltonian is used [see Fig.~\ref{fig:schematic}(c)]. Removing the constraint of traversing continuously over all the parameter space could provide a deeper insight on the Thouless pumping and could reduce the experimental complexity. The latter is relevant when experimental realization of the Hamiltonian for all the continuous parameters is inaccessible or problematic, e.g., due to hardware limitations. 

Our approach to realize Thouless pumping is to apply the Hamiltonian originally used in the conventional Thouless pumping, but with the use of only a number of discretely sampled parameter points, see Fig.~\ref{fig:schematic} (b) and (c). Consider the parameter $\phi$ that parameterizes the Hamiltonian $H(\phi)$, where $\phi=\omega t$ for conventional adiabatic approach. Consider $\phi \in [0, 2\pi]$ for the pumping in one cycle. Instead of varying $\phi$ continuously and smoothly over all the parameter range, we sample  $\phi= \phi_1, \phi_2, \ldots, \phi_P$ by $P$ equally spaced points along the adiabatic evolution path:
\begin{equation}\label{phip}
	\phi_{p}=\frac{2\pi}{2P}(2 p-1), ~~~~\text{for}~p=1,2,\cdots,P.
\end{equation}
See Fig.~\ref{fig:schematic}(c) for the control parameters used in our adiabatic approach. At each of the $p$th point ($p\in \{ 1,2,\cdots,P\}$), we apply $H(\phi_{p})$ for a time $t_{p}$ that in general depends on the value of $\phi=\phi_{p}$.

After our adiabatic jumping control in one pumping periodic $T$, the entire evolution operator is expressed as
\begin{equation}
	U_T = U(\phi_{P}) U(\phi_{P-1}) \cdots U(\phi_2) U(\phi_1),
\end{equation}
where the evolution operator driven by the Hamiltonian at a path point $\phi_{p}$ is given by 
\begin{equation}	U(\phi_{p}) = e^{-i H(\phi_{p}) t_{p}}.
\end{equation} 
By the use of Eq.~\eqref{eq:momentum_basis}, we have
\begin{equation}\label{eq:U_T_k}
	U_T = \sum_{k=-l}^{l} U_k(T) \otimes |k\rangle\langle k|, 
\end{equation}
where the evolution operator $U_k(t)$ for the pseudospin corresponding to the momentum state $|k\rangle$ is given by Eq.~\eqref{eq:U_k_General}. For our jumping scheme, it reads
\begin{align}\label{eq:Uk_jump}
	U_k(t)  = & e^{-i H_k(\phi_p) (t-T_p)} e^{-i H_k(\phi_{p-1})t_{p-1}} \cdots \\ \nonumber
	& \times 	e^{-i H_k(\phi_{2})t_{2}} e^{-i H_k(\phi_{1})t_{1}}, ~\text{if}~t\in [T_p, T_{p+1}),
\end{align}
with $T_p=\sum_{j=1}^{p-1}t_{j}$ being the time duration after applying the Hamiltonian $H(\phi)$ at the points $p=1,2,\cdots,\wzy{P}-1$. 

In the following, we determine the value of $t_{p}$ for Thouless pumping. According to the adiabatic jumping protocols for a single qubit in Refs.~\cite{Wang2016,Xu2019}, setting $t_{p}=\pi/(2\Omega_{\phi_{p}}^{(k)})$ \wzy{for a $\pi$ rotation of the pesudospin} with 
\begin{equation}
	2\Omega_{\phi_{p}}^{(k)}=2 \sqrt{(h_{x}^{(k)})^2+(h_{y}^{(k)})^2+(h_{z}^{(k)})^2}, \label{eq:Rabi}
\end{equation}
being the Rabi frequency for the pseudospin corresponding to the momentum state $|k\rangle$, all non-adiabatic transitions can be fully eliminated even in a short pumping period $T$ for the pseudospin corresponding to the momentum state $|k\rangle$, if the path parameter $\phi$ of the pseudospin is along a geodesic. Because the physical time $t_{p}$ to apply $H(\phi_{p})$ can only have a fixed value, our control does not fully eliminate all the non-adiabatic transitions simultaneously for all pseudospins with different values of $k$, which is different from the case of single qubit in Refs.~\cite{Wang2016,Xu2019}. As a consequence, in this work we use $t_{p}=\pi/(2\Omega_{\phi_{p}}^{(l)})$ to fully eliminate the non-adiabatic transitions for the pseudospin corresponding to the momentum state $|k=l\rangle$. (We note that other choices of $k$ are possible.) Then we choose the control parameters such that the evolution driven by $H_k(\phi_p)$ follows a geodesic for $k=l$, namely, we set $2\delta_0=\Delta_0$ in the parameters of $h_{\alpha}^{(l)}$ ($\alpha=x,y,z$). \wzy{Under this condition, we obtain a constant value of $t_{p}$, which we denote by $t_{\text{pulse}}=\pi/(2\Delta_{0})$}. For this control, the non-adiabatic transitions can still be suppressed for the pseudospins of other momentum states $|k\neq l\rangle$, even though these non-adiabatic transitions can not be fully eliminated, as we will show in Sec.~\ref{sec:Numerical}. \wzy{The error of adiabatic transport can be effectively averaged out by increasing the pulse number $P$ according to Ref.~\cite{Wang2016} (see Appendix~\ref{Robustnessofthequantization}).}
\wzy{However, since the total evolution time $T=P t_{\text{pulse}}$, in experiments the maximum number of pulses is limited by the coherence time of the quantum system.}

\section{Numerical results}\label{sec:Numerical}
We numerically compare the performance of Thouless pumping in momentum space for our jumping adiabatic scheme and the conventional adiabatic approach. 

\begin{figure}
	\centering
	\includegraphics{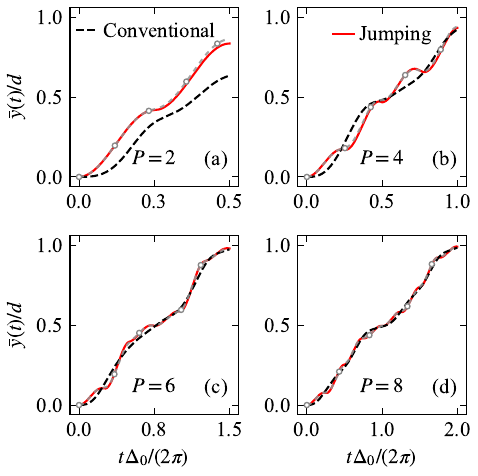}
	\caption{Comparison of $\bar{y}(t)$ between the conventional adiabatic scheme (black dashed lines) and the adiabatic jumping scheme (red lines) for different pumping period $T$. Here $T=P\pi/(2\Delta_{0})$. (a)-(d) Time evolution of $\bar{y}(t)$ for $P = 2$, $4$, $6$, and $8$, respectively.
		The parameters used in the simulation are: $J = -2\pi$, $\Delta_0 = 4\pi$, $\delta_0 = 0.5\Delta_0$ and lattice number $N = 801$. \wzy{The gray dashed line with open circles shows the results of the jumping scheme with sinusoidal transitions given by Eq.~(\ref{sinusoidaltransition}), with $\tau=0.5t_{\text{pulse}}$.}}
	\label{fig:berry_accumulation}
\end{figure}

In Fig.~\ref{fig:berry_accumulation}, we calculate time evolution of $\bar{y}(t)/d$. For the conventional adiabatic scheme, when the pumping period $T$ is large enough for a sufficiently slow continuous modulation of the Hamiltonian,  $\bar{y}(T)/d$ approaches the quantized value $C=1$. For our jumping scheme, despite of the fast changes of control parameters, $\bar{y}(T)/d$ also approaches the quantized value $C=1$ when the number $P$ increases. We note that when the pumping period $T$ is short [see Fig.~\ref{fig:berry_accumulation}(a)], our jumping scheme performs better than the conventional Thouless pumping. This indicates that the adiabatic jumping scheme can achieve a higher fidelity of  adiabatic evolution than the conventional method. 

\begin{figure}
	\includegraphics[width=0.9\columnwidth]{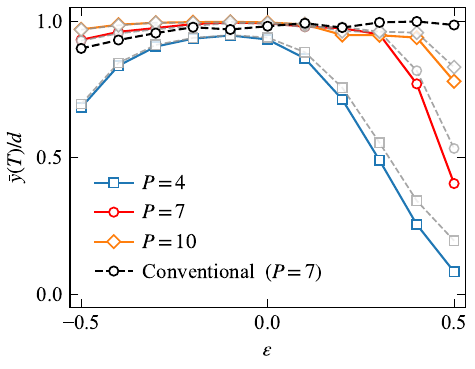}
	\caption{Robustness of adiabatic jumping approach in the presence of error $\varepsilon$. $\bar{y}(T)/d$ remains close to the quantized value $C=1$ for a wider range of $\varepsilon$ when $P$ is larger. Comparing the conventional scheme (black circles) with the jumping approach with the same pumping period $T$ (red circles), the jumping approach shows better performance than the conventional scheme for $\varepsilon<0$. The parameters used in the simulation are: $J = -2\pi$, $\Delta_0 = 4\pi$, $\delta_0 = 0.5\Delta_0$ and lattice number $N = 801$. \wzy{The gray dashed lines are the results when the discrete jumps of $\phi$ are changed to sinusoidal transitions given by Eq.~(\ref{sinusoidaltransition}) using a transition time $\tau=0.5t_{\text{pulse}}$.}}
	\label{fig:robustness}
\end{figure}

Because of its adiabatic nature, our jumping scheme is robust to error in the pumping parameters. In Fig.~\ref{fig:robustness}, we introduce an amplitude error to the Hamiltonian as $H\rightarrow (1 + \varepsilon) H$, where $\varepsilon$ denotes the strength of the error. The quantity $\bar{y}(T)/d$ of the adiabatic jumping scheme maintains a value close to $C=1$ even for relatively large errors, demonstrating the superior robustness of our approach. The robustness is stronger when $P$ and hence $T$ are larger. 
\wzy{We note that the conventional scheme achieves better adiabatic transport as $\varepsilon$ increases, due to a larger energy gap (which is proportional to $1+\varepsilon$). In contrast, for our scheme, an error $\varepsilon\neq 0$ alters the $\pi$ rotation of the pseudospin [see Eq.~\eqref{eq:Rabi}], thereby degrading the transport performance. The asymmetry in performance between $\varepsilon> 0$ and $\varepsilon< 0$ stems from the asymmetry of the Rabi frequency $\Omega^{(k)}_{\phi_p}$ with respect to the momentum states $|k\rangle$.}


Then we demonstrate the quantized transport in real space by simulating the evolution of a Wannier state under the adiabatic jumping approach. The mean displacement of the wave packet after one pumping cycle serves as a direct measure of the topological pumping.

We initialize the system in a Wannier state localized at a specific unit cell $m =m_0$. The Wannier state is constructed as a coherent superposition of Bloch states from the lower band:
\begin{equation}
	|\Psi_{-,m_0}(0)\rangle = \sum_{k=-l}^{l} \psi_k |s_k(0)\rangle \otimes |k\rangle,
\end{equation}
where $\psi_k = \frac{1}{\sqrt{N}} e^{-i\frac{2\pi}{N} k m_0}$ and the initial state $|s_k(0)\rangle$ of the pseudospin of each $k$ is the ground state of the initial $H_k$ at $t=0$ (see Appendix~\ref{app:displacement} for the expression of $|s_k(0)\rangle$).
Using the results in Sec.~\ref{sec:Theory}, the Wannier state evolves as
\begin{equation}
	|\Psi_{-,m_{0}}(t)\rangle = \sum_{k} \psi_k |s_k(t)\rangle \otimes |k\rangle,
\end{equation}
with $|s_k(t)\rangle = U_k(t)|s_k(0)\rangle$, where $U_k(t)$ is the time-evolution operator in $k$-space [see Eq.~\ref{eq:U_k_General} for the general control and Eq.~\ref{eq:Uk_jump} for the special case of the jumping scheme]. In Appendix~\ref{app:displacement}, we obtain the mean displacement at any time $t$ in position space
\begin{equation}
	\Delta x(t) = \langle \hat{x}(t) \rangle - \langle \hat{x}(0) \rangle,
\end{equation}
where the expectation value of the position is given by 
\begin{equation}
	\langle \hat{x}(t) \rangle = \sum_{k\neq k'} \frac{-i}{2} \frac{(-1)^{k'-k}}{\sin\frac{\pi(k'-k)}{N}} \psi_k^* \psi_{k'} \langle s_k(t)|s_{k'}(t)\rangle.
\end{equation}

\begin{figure}
	\centering
	\includegraphics[width=1\columnwidth]{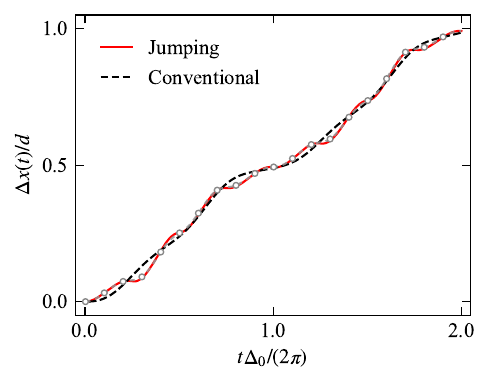}
	\caption{Mean displacement $\Delta x(t)$ during one pumping cycle for the adiabatic jumping approach with $P=8$ (red solid line). The black dashed line is the result for the conventional Thouless pumping using the same pumping period $T=P\pi/(2\Delta_{0})$. The result shows a quantized displacement of one unit cell ($\Delta x(T)/d = 1$) at the end of one pumping period $T$, confirming the successful topological pumping. We choose $J = -2\pi$, $\Delta_0 = 4\pi$, $\delta_0 = 0.5\Delta_0$, lattice number $N = 801$, and Wannier state is initially localized at $m_0 = 0$ in the position space. \wzy{The gray dashed line with open circles shows the results of the adiabatic jumping scheme with sinusoidal transitions defined in Eq.~(\ref{sinusoidaltransition}), with transition time $\tau=0.5t_{\text{pulse}}$.}}
	\label{fig:displacement}
\end{figure}

In Fig.~\ref{fig:displacement} we presents the mean displacement $\Delta x(t)$ during one pumping cycle.  The result clearly shows a quantized displacement of one unit cell, demonstrating that the adiabatic jumping approach successfully achieves topological pumping. 

\begin{figure}
	\centering
	\includegraphics[width=1\columnwidth]{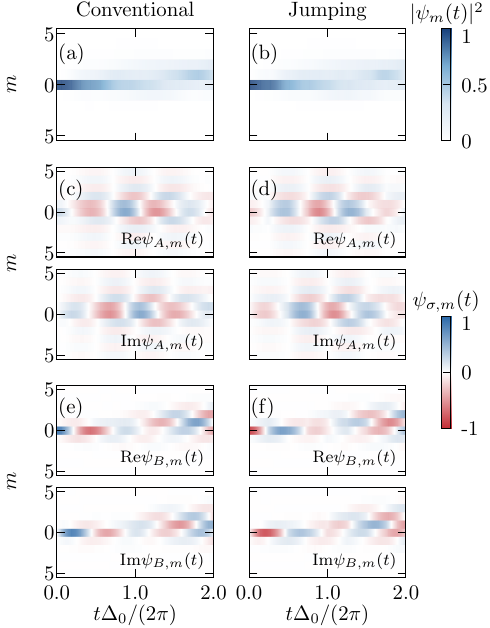}
	\caption{\wzy{The temporal evolution of the wave function during one pumping period $T$. (a) and (b) show the evolution of the probability $|\psi_m(t)|^2$ for the conventional and jumping schemes, respectively. (c) and (e) show the real and imaginary parts of $\psi_{\sigma,m}(t)$ for $\sigma=A$ and $B$, respectively, for the conventional scheme. (d) and (f) are the same as (c) and (e), but for the jumping scheme. Here $J = -2\pi$, $\Delta_0 = 4\pi$, $\delta_0 = 0.5\Delta_0$, $P=8$, and lattice number $N = 801$. The Wannier state is initially localized at $m_0 = 0$ in real space.}}
	\label{fig:waveFunction}
\end{figure}

\wzy{In Fig.~\ref{fig:waveFunction}, we show the simulated temporal evolution of the wave function over a single pump cycle, along with the resulting probability distribution $|\psi_m(t)|^2 = \sum_{\sigma=A,B}|\psi_{\sigma,m}(t)|^2$ at each unit cell $m$. Here, the projected amplitudes are defined as $\psi_{\sigma,m}(t)=\langle \sigma,m|\Psi_{-,m_0}(t)\rangle$, where $\sigma=A,B$ denotes the site label and $m$ is the unit cell index in real space. By comparing these dynamics with those of the conventional scheme, we further confirm that the jumping scheme can realize topological quantum transport.}

\wzy{To demonstrate the robustness of the jumping scheme on the variations in $\phi$, we also consider simulations where the stepwise changes in Eq.~(\ref{phip}) are replaced by a sinusoidal transition from $\phi_p$ to $\phi_{p+1}$ for every $p=1,2,\dots,P-1$ (see Fig.~\ref{fig:schematic}). Specifically, for the jumping scheme with a smooth transition, we modify the change from $\phi_p$ to $\phi_{p+1}$ according to 
\begin{align}\label{sinusoidaltransition}
		\phi=\frac{2\pi}{2P}(2p-1)+\frac{2\pi}{2P}\{1+\sin[\pi(t-pt_{\text{pulse}})/\tau]\},
\end{align}
when the evolution time $t$ satisfies $|t-pt_{\text{pulse}}|\leq\tau/2$, where $\tau$ denotes the transition time. Comparing the results in Figs.~\ref{fig:berry_accumulation} to \ref{fig:displacement} obtained from instantaneous and smooth changes of $\phi$ shows that the jumping scheme is robust to variations in $\phi$.}

\section{CONCLUSION AND DISCUSSION}\label{sec:Conclusion}
In summary, we have shown that topological Thouless pumping can be adiabatically realized using only discrete values of the Hamiltonian, contrary to the conventional adiabatic approach that uses continuous and slow change of the Hamiltonian. Based on the necessary and sufficient condition for quantum adiabatic evolution, we can effectively suppress nonadiabatic transitions while rapidly varying the Hamiltonian. More importantly, we can drive the evolution of quantum states using a discrete Hamiltonian form, thus making it possible to bypass some experimentally inaccessible regimes. Compared with the STA scheme, our work does not require the introduction of additional complicated auxiliary fields. This work based on adiabatic jumping still exhibits excellent robustness even in the presence of large control-field errors. This extends the application of the necessary and sufficient condition for quantum adiabatic evolution to interacting quantum systems and provides a new perspective for adiabatic control in condensed matter physics.

\section*{Acknowledgements}
We acknowledge supports from National Natural Science Foundation of China (Grant No. 12575017), Natural Science Foundation of Guangdong Province (Grant No. 2025A1515011571), and Guangdong Provincial Quantum Science Strategic Initiative (Grant No. GDZX2505006, GDZX2203001). L.Y. and Y.W. contributed equally to this work.

\appendix

\section{Calculation of displacement}\label{app:displacement}

The position operator in real space is defined as
\begin{equation}
	\hat{x} = \sum_{m=-l}^{l} m |m\rangle\langle m|.
\end{equation}
where $m$ depicts the positions of the unit cells and $l=(N-1)/2$ with $N$ being the total number of unit cells. Using Eq.~\eqref{eq:Fourier_transform}, we write the position operator $\hat{x}$ in terms of the momentum basis $|k\rangle$ as
\begin{equation}
	\hat{x} = \frac{1}{N} \sum_{m=-l}^{l} m \sum_{k'=-l}^{l} \sum_{k=-l}^{l} e^{i\frac{2\pi }{N}m(k'-k)} |k\rangle\langle k'|.
\end{equation}
Using 
\begin{align}
	\sum_{m=-l}^{l} m e^{i\theta m} & = \frac{\partial}{\partial\alpha} \left(\sum_{m=-l}^{l} e^{i\theta m \alpha}\right)\Big|_{\alpha=1}  \\ \nonumber
	& = \frac{1}{i\theta} \frac{\partial}{\partial\alpha} \frac{\sin\frac{\theta}{2}\alpha(2l+1)}{\sin\frac{\theta}{2}\alpha}\Big|_{\alpha=1},
\end{align}
we obtain
\begin{equation}
	\hat{x} = \sum_{k'\neq k} \frac{-i}{2} \frac{(-1)^{k'-k}}{\sin\frac{\pi(k'-k)}{N}} |k\rangle\langle k'|.
\end{equation}

To find the mean displacement $\Delta x(t)$ for the Wannier state, we first find the expression of the initial ground state $|s_k(0)\rangle$ of the Hamiltonian $H_k$ at $t=0$. Using the parameters $h_x^{(k)}= J(1 + \cos K_k)$, $h_y^{(k)} = J \sin K_k$, and $h_z^{(k)} = \Delta_0$ at $\phi = 0$ with $K_k = 2\pi k/N$, we have 
\begin{equation}
	|s_k(0)\rangle = \sin\frac{\theta_k}{2} |A\rangle - e^{i\phi_k} \cos\frac{\theta_k}{2} |B\rangle,
\end{equation}
where $\sin\frac{\theta_k}{2} = \sqrt{\frac{\Omega_k - \Delta_0}{2\Omega_k}}$, $\cos\frac{\theta_k}{2} = \sqrt{\frac{\Omega_k + \Delta_0}{2\Omega_k}}$, and $e^{i\phi_k} = e^{iK_k/2}$, with $\Omega_k = \sqrt{[2J\cos(K_k/2)]^2 + \Delta_0^2}$.

Using the $k$-space representation of $\hat{x}$ and the evolved state $|\Psi_{-,m_{0}}(t)\rangle = \sum_{k} \psi_k |s_k(t)\rangle \otimes |k\rangle$, the position expectation value is
\begin{equation}
	\langle \hat{x}(t) \rangle = \sum_{k\neq k'} \frac{-i}{2} \frac{(-1)^{k'-k}}{\sin\frac{\pi(k'-k)}{N}} \psi_k^* \psi_{k'} \langle s_k(t)|s_{k'}(t)\rangle,
\end{equation}
where $|{s_{k}(t)} \rangle=U_{k}(t)|s_{k}(0)\rangle$. Here $U_{k}(t)$ is given by Eq.~\ref{eq:U_k_General} for the general control and Eq.~\ref{eq:Uk_jump} for the special case of the jumping scheme.

\wzy{\section{Adiabaticity of the jumping scheme}}\label{Robustnessofthequantization}
Here we provide theoretical details to show that the jumping scheme achieve better quantum adiabatic evolution when the number $P$ of the sampling points $\phi_1, \phi_2,\ldots, \phi_P$ for the parameter $\phi$ increases. According to Eq.~\eqref{eq:U_T_k}, the quantum evolution is decoupled for different momentum state $|k\rangle$. This allows us to analyze the quantum dynamics $U_k$ in Eq.~\eqref{eq:U_T_k} within a given momentum $k$. According to Ref.~\cite{Wang2016}, the evolution operator $U_k$ can be decomposed into $U_k=U_{\mathrm{adia},k} U_{\mathrm{Dia}}$, where $U_{\mathrm{adia},k}$ denotes the ideal adiabatic evolution operator and $U_{\mathrm{Dia}}$ represents the nonadiabatic correction. For simplicity, we omit the $k$ dependence in the notation of $U_{\mathrm{Dia}}$ and other quantities in this appendix. 

According to Ref.~\cite{Wang2016}, the nonadiabatic correction is given by
\begin{align}
	U_{\mathrm{Dia}}=\mathcal{T}e^{\left[i\int_0^t\left(F_{+,-}(t')G_{+,-}(t')+\mathrm{H.c.}\right)dt'\right]},
\end{align}
where $\mathcal{T}$ denotes the time-ordering operator. Here, $F_{+,-}(t)=e^{i\int_{0}^{t}2\Omega(t') dt'}$, where the integral $\int_{0}^{t}2\Omega(t') dt'$ represents the relative dynamical phase between the two eigenstates. $G_{+,-}(t)$ is a geometric function
\begin{align}
	G_{+,-}(t)=ie^{i\gamma(t)}\langle s_{+}(t)|\partial_t| s_{-}(t)\rangle| s_{+}(0)\rangle\langle s_{-}(0)|,	
	\label{eq:G_k}
\end{align}
where $\gamma(t)=\gamma_{-}(t)-\gamma_{+}(t)$ is the relative geometric phase, and $|s_{\pm}(t)\rangle$ are the instantaneous eigenstates of Hamiltonian.

The deviation from the adiabatic evolution is described by $\|D_{\mathrm{Dia}}\|\equiv \|U_{\mathrm{Dia}}-I\|$, where $\|\cdot\|$ denotes the spectral norm. According to Eq.~(18) in Ref.~\cite{Wang2016} 
\begin{align}\label{Dia}
	\|D_{\mathrm{Dia}}\| <&
	\sqrt{\xi}\left(g_{\mathrm{tot}}^2+w_{\mathrm{tot}}\right)
	(\phi-\phi_0)^2 \nonumber\\
	&+ \left(\sqrt{\xi}+\xi\right)g_{\mathrm{tot}},
\end{align}
where $g_{\mathrm{tot}}=\sum_{\alpha\neq\beta}\sup\|G_{\alpha,\beta}(\phi)\|$ and
$w_{\mathrm{tot}}=\sum_{\alpha\neq\beta}\sup\left\|\frac{d}{d\phi}G_{\alpha,\beta}(\phi)\right\|$ with
$\alpha,\beta\in\{+,-\}$. Since the geometric path has finite length, both $g_{\mathrm{tot}}$ and $w_{\mathrm{tot}}$ are bounded independently of $P$. $\xi$ is the upper bound of $\left|\int_0^t F_{\pm,\mp}(t')dt'\right|$ (see Ref.~\cite{Wang2016}). In the following, we will derive that 
\begin{align}
	\xi=\left[2c+4c^2\frac{\pi}{\Delta_{0}}\left(\sqrt{4J^2+\Delta_0^2}-|J|\right)\right]
	\frac{2\pi}{P},\label{eq:xi_k_simple}
\end{align}
which shows that $\xi\to 0$ in the limit of $P\to\infty$ and hence the quantum evolution is adiabatic according to Eq.~\eqref{Dia}.

Explicitly, using $D\equiv4J^{2}\cos^{2}\frac{K_{k}}{2}+\Delta_{0}^{2}\sin^{2}\frac{K_{k}}{2}>0$ and $J<\Delta_0$, we obtain the following upper bound for $g_{\mathrm{tot}}:$
\begin{align}\label{Gk}
	g_{\mathrm{tot}}\le\left|\frac{\Delta_0}{J}\right|.
\end{align}
Similarly, the bounds $1-\cos^2\frac{K_k}{2}\sin^2t\le1$, $\sin^2t\cos^2t\le\frac14$, and
$\Omega^2(t)\ge D$ yield
\begin{align}\label{dGk}
	w_{\mathrm{tot}}
	\le\frac{|\Delta_0|}{2J^2}
	\left[J^2+\frac{\Delta_0^2(4J^2+\Delta_0^2)}{16J^2}\right]^{1/2}.
\end{align}

The derivation of Eq.~(\ref{eq:xi_k_simple}) is given below. According to Ref.~\cite{Wang2016}, on the interval $(T_{p-1},T_p]$, we have $\int_{0}^{t}F_{+,-}(t') dt'=\sum_{j=1}^{p}F_{j}\frac{2\pi}{P}$, with $F_{+,-}(\phi_j)\equiv F_{j}$. Using $F_{j}=e^{i\varphi_{j}}F_{j-1}$, with $\varphi_{j}=2\Omega_{\phi_j}t_j$, we obtain 
\begin{align}\label{Fj+1-Fj}
	F_{j}=a_{j}(F_{j}-F_{j-1}),
\end{align}
where $a_{j}=1/(1-e^{-i\varphi_{j}})$.
Because $\Omega_{\phi_j}\le\Omega^{(l)}$, and $2\Omega^{(l)}t_j=\pi$, we know that $\frac{\varphi_{j}}{2}\in (0,\frac{\pi}{2}]$. Since $|a_{j}|$ decreases monotonically with $\varphi_{j}$, its maximum is attained at the minimum value of $\varphi_{j}$. Because $\varphi_{j}=(\Omega_{\phi_j}/\Omega^{(l)})\pi\ge\frac{|J|}{\sqrt{4J^2+\Delta_0^2}}\pi$, thus
\begin{align}
	|a_{j}|\le\left|\frac{1}{2\sin [|J|/(2\sqrt{4J^2+\Delta_0^2})\pi]}\right|\equiv c.\label{ak}
\end{align}
Substituting Eq.~(\ref{Fj+1-Fj}) into $\sum_{j=1}^{p}F_{j}$ and rearranging gives
\begin{align}\label{eq:xi_k}
	\left|\sum_{j=1}^{p}F_{j}\right|=\left|a_{p}F_{p}+(1-a_{2})F_{1}-\sum_{j=2}^{p-1}F_{j}(a_{j+1}-a_{j})\right|.
\end{align}
First, using Eq.~(\ref{ak}) and $|F_{j}|=1$, we obtain the bound for the first two terms $|a_{p}F_{p}+(1-a_{2})F_{1}|\le2c$. According to Eq.~(\ref{ak}) and $t_j=\pi/(2\Delta_0)$,  we obtain
\begin{align}\label{Fajk}
	\left|\sum_{j=2}^{p-1}F_{j}(a_{j+1}-a_{j})\right|&\le c^2\sum_{j=2}^{p-1}\left|e^{-i\varphi_{j+1}}-e^{-i\varphi_{j}}\right|\nonumber\\ &\le c^2 \sum_{j=2}^{p-1}\left|\varphi_{j+1}-\varphi_{j}\right|\\
	&\leq 4c^2\frac{\pi}{\Delta_{0}}\Bigg(\sqrt{4J^2+\Delta_0^2}-|J|\Bigg)\nonumber.
\end{align}
It is obvious that the same bound applies to $\int_{0}^{t}F_{-,+}(t') dt'$. Combining these results gives Eq.~(\ref{eq:xi_k_simple}), demonstrating that increasing the pulse number $P$ suppresses the nonadiabatic effects.


\begin{thebibliography}{99}
	\bibitem{Thouless1983} D. J. Thouless, \textit{Quantization of particle transport}, Phys. Rev. B \textbf{27}, 6083 (1983).
	\bibitem{Niu1984} Q. Niu and D. J. Thouless, \textit{Quantised adiabatic charge transport in the presence of substrate disorder and many-body interaction}, J. Phys. A \textbf{17}, 2453 (1984).
	\bibitem{Xiao2010} D. Xiao, M.-C. Chang, and Q. Niu, \textit{Berry phase effects on electronic properties}, Rev. Mod. Phys. \textbf{82}, 1959 (2010).
	\bibitem{Citro2023} R. Citro and M. Aidelsburger, \textit{Thouless pumping and topology}, Nat. Rev. Phys. \textbf{5}, 87 (2023).
	\bibitem{Nakajima2016} S. Nakajima, T. Tomita, S. Taie, T. Ichinose, H. Ozawa, L. Wang, M. Troyer, and Y. Takahashi, \textit{Topological Thouless pumping of ultracold fermions}, Nat. Phys. \textbf{12}, 296 (2016).
	\bibitem{Lohse2016} M. Lohse, C. Schweizer, O. Zilberberg, M. Aidelsburger, and I. Bloch, \textit{A Thouless quantum pump with ultracold bosonic atoms in an optical superlattice}, Nat. Phys. \textbf{12}, 350 (2016).
	\bibitem{Lohse2018} M. Lohse, C. Schweizer, H. M. Price, O. Zilberberg, and I. Bloch, \textit{Exploring 4D quantum Hall physics with a 2D topological charge pump}, Nature (London) \textbf{553}, 55 (2018).
	\bibitem{Benalcazar2022} W. A. Benalcazar, T. L. Hughes, and B. A. Bernevig, \textit{Higher-order topological pumping and its observation in photonic lattices}, Phys. Rev. B \textbf{105}, 195129 (2022).
	\bibitem{Kraus2012} Y. E. Kraus, Y. Lahini, Z. Ringel, M. Verbin, and O. Zilberberg, \textit{Topological states and adiabatic pumping in quasicrystals}, Phys. Rev. Lett. \textbf{109}, 106402 (2012).
	\bibitem{Zilberberg2018} O. Zilberberg, S. Huang, J. Guglielmon, M. Wang, K. P. Chen, Y. E. Kraus, and M. C. Rechtsman, \textit{Photonic topological boundary pumping as a probe of 4D quantum Hall physics}, Nature (London) \textbf{553}, 59 (2018).
	\bibitem{Cerjan2020} A. Cerjan, M. Wang, S. Huang, K. P. Chen, and M. C. Rechtsman, \textit{Thouless pumping in disordered photonic systems}, Light Sci. Appl. \textbf{9}, 178 (2020).
	\bibitem{Messiah1965} A. Messiah, \textit{Quantum Mechanics} (North-Holland, Amsterdam, 1965), Vol. II.
	\bibitem{Privitera2018} L. Privitera, A. Russomanno, R. Citro, and G. E. Santoro, \textit{Nonadiabatic breaking of topological pumping}, Phys. Rev. Lett. \textbf{120}, 106601 (2018).
	\bibitem{Arceci2020} L. Arceci, S. Barbarino, R. Fazio, and G. E. Santoro, \textit{Dissipation assisted Thouless pumping in the Rice-Mele model}, J. Stat. Mech. (2020) 043101.
	\bibitem{Fedorova2020} Z. Fedorova, M. Lohse, C. Schweizer, M. Aidelsburger, and I. Bloch, \textit{Observation of topological transport quantization by dissipation in fast Thouless pumps}, Nat. Commun. \textbf{11}, 3758 (2020).
	\bibitem{Schouten2021} K. J. M. Schouten and V. Cheianov, \textit{Rapid-cycle Thouless pumping in a one-dimensional optical lattice}, Phys. Rev. A \textbf{104}, 063315 (2021).
	\bibitem{Malikis2022} S. Malikis and V. Cheianov, \textit{An ideal rapid-cycle Thouless pump}, SciPost Phys. \textbf{12}, 203 (2022).
	\bibitem{Liu2025} W. Liu, Y. Ke, and C. Lee, \textit{Shortcuts to adiabatic Thouless pumping}, Phys. Rev. A \textbf{112}, 013317 (2025).
	\bibitem{Berry2009} M. V. Berry, \textit{Transitionless quantum driving}, J. Phys. A \textbf{42}, 365303 (2009).
	\bibitem{Demirplak2005} M. Demirplak and S. A. Rice, \textit{Assisted adiabatic passage revisited}, J. Phys. Chem. B \textbf{109}, 6838 (2005).
	\bibitem{GueryOdelin2019} D. Gu{\'e}ry-Odelin, A. Ruschhaupt, A. Kiely, E. Torrontegui, S. Mart{\'i}nez-Garaot, and J. G. Muga, \textit{Shortcuts to adiabaticity: Concepts, methods, and applications}, Rev. Mod. Phys. \textbf{91}, 045001 (2019).
	\bibitem{An2016} S. An, D. Lv, A. del Campo, and K. Kim, \textit{Shortcuts to adiabaticity by counterdiabatic driving for trapped-ion displacement in phase space}, Nat. Commun. \textbf{7}, 12999 (2016).
	\bibitem{Chen2010a} X. Chen, A. Ruschhaupt, S. Schmidt, A. del Campo, D. Gu{\'e}ry-Odelin, and J. G. Muga, \textit{Fast optimal frictionless atom cooling in harmonic traps: Shortcut to adiabaticity}, Phys. Rev. Lett. \textbf{104}, 063002 (2010).
	\bibitem{Chen2010b} X. Chen, I. Lizuain, A. Ruschhaupt, D. Gu{\'e}ry-Odelin, and J. G. Muga, \textit{Shortcut to adiabatic passage in two- and three-level atoms}, Phys. Rev. Lett. \textbf{105}, 123003 (2010).
	\bibitem{Wang2016} Z.-Y. Wang and M. B. Plenio, \textit{Necessary and sufficient condition for quantum adiabatic evolution by unitary control fields}, Phys. Rev. A \textbf{93}, 052107 (2016).
	\bibitem{Xu2019} K. Xu, Z.-Y. Wang, M. B. Plenio, and J. Zhang, \textit{Breaking the quantum adiabatic speed limit by jumping along geodesics}, Sci. Adv. \textbf{5}, eaax3800 (2019).
	\bibitem{Chen2024} C. Chen, J.-Y. Lu, X.-Y. Chen, and Z.-Y. Wang, \textit{Fast adiabatic preparation of multisqueezed states by jumping along the path}, Phys. Rev. A \textbf{110}, 012601 (2024).
	\bibitem{Zhang2025} J. Zhang, T. Xing, and G. Long, \textit{Speeding up adiabatic holonomic quantum gates via $\pi$-pulse modulation}, Fundam. Res. \textbf{5}, (2025).
	\bibitem{Zheng2022} W. Zheng, Y. Zhang, X. Song, and Y. Wang, \textit{Accelerated Quantum Adiabatic Transfer in Superconducting Qubits}, Phys. Rev. Appl. \textbf{18}, 044014 (2022).
	\bibitem{Gong2023} M. Gong, Y. Wu, S. Wang, and C. Zhang, \textit{Accelerated quantum control in a three-level system by jumping along the geodesics}, Phys. Rev. A \textbf{107}, L040602 (2023).
	\bibitem{Xing2026} T. Xing, J. Zhang, and G. Long, \textit{Accelerating quantum adiabatic evolution with $\pi$-pulse sequences}, Sci. China-Phys. Mech. Astron. \textbf{69} (3), (2026).
	\bibitem{LiuWang2022} Y. Liu and Z.-Y. Wang, \textit{Shortcuts to Adiabaticity with Inherent Robustness and without Auxiliary Control}, arXiv:2211.02543 (2022).
	\bibitem{RiceMele1982} M. J. Rice and E. J. Mele, \textit{Elementary excitations of a linearly conjugated diatomic polymer}, Phys. Rev. Lett. \textbf{49}, 1455 (1982).
	\bibitem{Asboth2016} J. K. Asb{\'o}th, L. Oroszl{\'a}ny, and A. P{\'a}lyi, \textit{A Short Course on Topological Insulators: Band Structure and Edge States in One and Two Dimensions} (Springer, Cham, 2016).
	
	
\end{thebibliography}
\end{document}